\documentclass[a4paper,10pt]{article}
\usepackage[utf8]{inputenc}
\usepackage{graphicx}
\usepackage[margin=2cm]{geometry}
\usepackage[hidelinks]{hyperref}

\newcommand{\nm}{\ensuremath{\,\mathrm{nm}}}
\newcommand{\um}{\ensuremath{\,\mathrm{\mu m}}}
\newcommand{\mm}{\ensuremath{\,\mathrm{mm}}}

\newcommand{\degrees}{\ensuremath{^\circ}}
\newcommand{\rfig}[1]{Figure~\ref{#1}}
\newcommand{\rsec}[1]{Section~\ref{#1}}

\title{The OpenFlexure Block Stage: Sub-100 nm fibre alignment with a monolithic plastic flexure stage}
\author{Qingxin Meng$^{1}$, Kerrianne Harrington$^{1,2}$, Julian Stirling$^{1}$, and Richard Bowman$^{1,}\footnote{Email: {\tt r.w.bowman@bath.ac.uk}}$\\
\small $^1$ Department of Physics, University of Bath, Claverton Down, Bath, BA2 7AY, UK\\
\small $^2$ Current address: Optoelectronics Research Centre, University of Southampton,\\\small  University Road, Southampton, SO17 1BJ, UK}
\date{}

\begin{document}

\maketitle

\begin{abstract}
 As 3D printers become more widely available, researchers are able to rapidly produce components that may have previously taken weeks to have machined. The resulting plastic components, having high surface roughness, are often not suitable for high-precision optomechanics. However, by playing to the strengths of 3D printing---namely the ability to print complex internal geometries---it is possible to design monolithic mechanisms that do not rely on tight integration of high-precision parts. Here we present a motorised monolithic 3D-printed plastic flexure stage with sub-100 nm resolution, that can perform automated optical fibre alignment.
\end{abstract}

\section{Introduction}
Good mechanical positioning is a critical factor in the performance of most optical instruments, but is often the most overlooked aspect of their design.  The widespread availability of high quality modular optomechanical components makes it simple to construct a new beamline with parts from stock, which is ideal for prototyping new systems.  However, this usually leads to a design that is large, costly, and heavy. With 3D printers becoming more widely available, the barrier to converting a digital design into a physical object is now so low that making a component can be both cheaper and quicker than purchasing one\cite{silver2019five}; although this is only true if well-tested designs are available. Parametric computer aided design (CAD) allows simple customisation of parts for increased versatility. This means designs can be combined and adapted, resulting in smaller, lighter, cheaper, and more tightly integrated components.

Material performance, engineering tolerance, and surface finish are major drawbacks of 3D printing for optomechanical instrumentation. The vast majority of 3D printers available in labs and workshops extrude thermoplastics, based on the RepRap design\cite{RepRap_bowyer_2011}. Commonly used thermoplastics have an elastic modulus over an order of magnitude lower than metals traditionally used for optomechanical components\cite{kamthai2015thermal,CRC_MechProps}. This results in lower stiffness components that have less resistance to vibration, though the increased damping of thermoplastics can counteract this\cite{riddell1966fatigue}. Poor engineering tolerance and surface finish of 3D printed parts can result in poor performance if a design relies on sliding elements such as dovetails.

\begin{figure}[b!]
\centering
 \includegraphics[width=.8\textwidth]{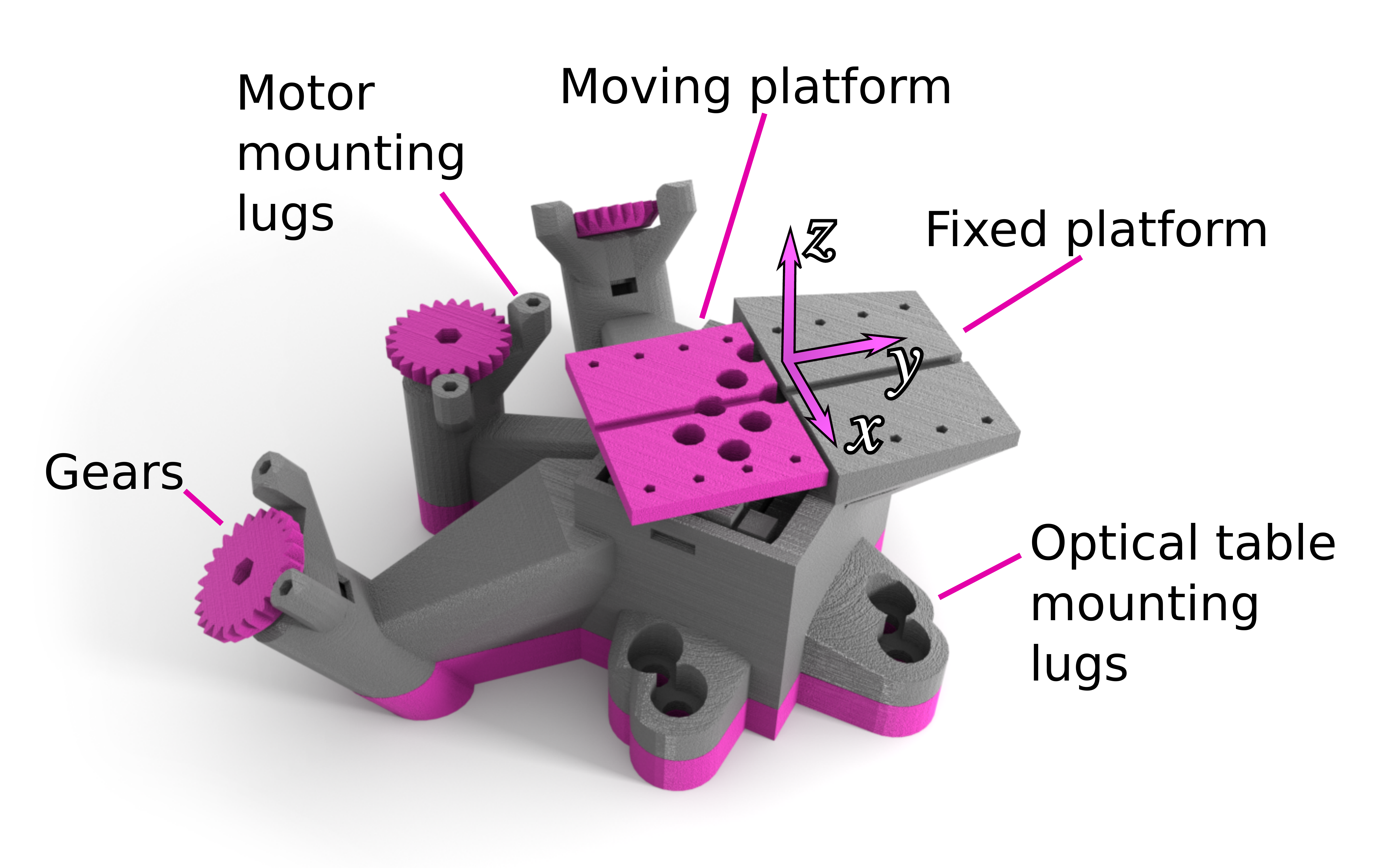}
 \caption{The OpenFlexure Block Stage. The mechanism is a single 3D printed part (grey), other parts (pink) such as the the base and moving platform are printed separately.
 The each axis of the stage is actuated by an M3 lead screw driven by a gear (pink). These gears can be driven by a smaller 3D printed gear (not shown) attached to a 28BYJ-48 stepper motors that can be mounted to the casing of the stage. }
 \label{Fig:Stage}
\end{figure}

It is possible to mitigate against these disadvantages by exploiting the advantages of 3D printing, resulting in instruments which can have extremely good performance.  The ability to form intricate geometries in a single part means that, as well as saving time and effort constructing the assembly, slippage and misalignment from attaching multiple parts is eliminated.  Similarly, the assembly can be smaller and more tightly integrated, as there is no need to allow access during construction. These small plastic mechanisms can be as stiff as, or stiffer than, large metal ones thanks to stiffness being inversely proportional to the cube of the length of mechanical linkages. By employing flexure hinges rather than separate sliding elements, a component with multiple degrees of freedom can be produced as a single part. The low elastic modulus of plastics allows a greater range of motion for a given lever length, thus allowing further size reductions and hence increasing stiffness. These monolithic components, being printed from a single material, can also be designed so much of the thermal expansion is cancelled out.

In this manuscript we introduce the OpenFlexure Block Stage\cite{BlockstageRepo} a motorised 3-axis translation stage with sub-$100\nm$ resolution. The parametric CAD design of the stage is provided so that it can be customised for different applications. The OpenFlexure Block Stage is not only a proof of principle design, but also a useful translation stage for common lab experiments, and a reusable component suitable for integration into more complicated instruments. We assess the mechanical performance of the stage, and demonstrate its use as an automated fibre alignment stage.

\begin{figure}[t!]
  \centering
  \includegraphics[width=.9\textwidth]{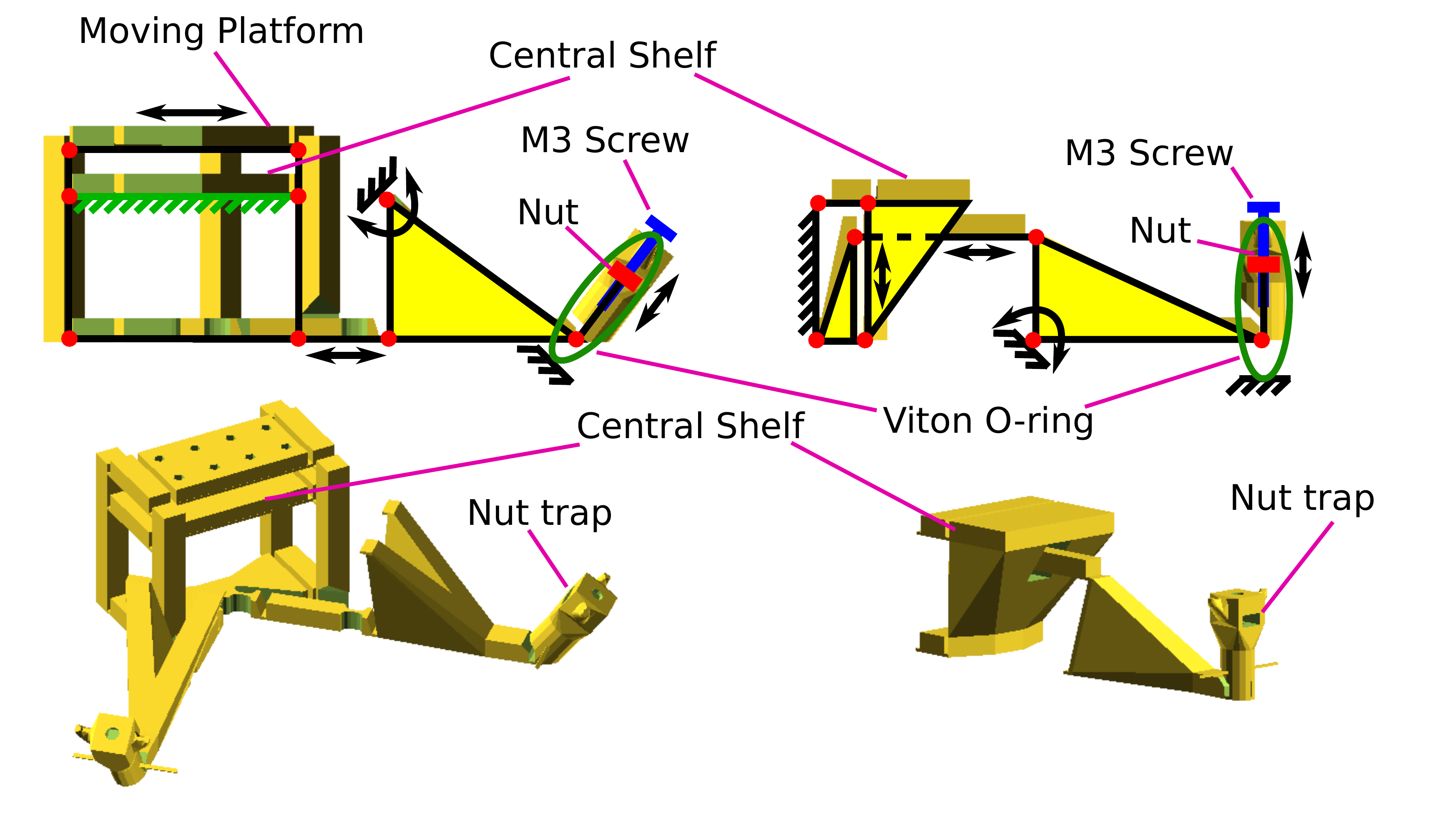}
 \caption{ \textit{Top left:} schematic of the $x$ and $y$-mechanisms. Red dots represent the position of flexure hinges, and black hatches represent attachment to the fixed casing of the stage. The green hatched portion represents the central shelf that is fixed during $xy$-motion but moves during $z$-travel. \textit{Top right:} schematic of $z$ mechanism. \textit{Bottom left and right:} cutaways of only the $xy$-mechanism and the $z$-mechanism respectively. The central shelf (labelled) is common to both cutaways. }\label{fig:mechanism}
\end{figure}

\section{ Mechanical design}
The OpenFlexure Block Stage is a monolithic 3-axis cartesian flexure stage, where all three actuators remain static and a single platform moves (See \rfig{Fig:Stage}).  The range of motion is approximately $2\times2\times2\mm$, actuated by three M3 screws. Each actuator geared down $5:1$ by a system of levers to give $100\um$ of stage motion per revolution of the M3 lead screws.  In order to be immediately useful for fibre alignment, it can be configured with both a fixed and a moving platform. Both platforms feature a central alignment groove and M3 screw holes, to allow the use of the same optics mounts used with commercial fibre launch platforms.

The OpenFlexure Block Stage draws on the OpenFlexure Microscope stage design\cite{sharkey2016one,ofm2019} for elements such as the ``table'' shaped moving stage.  The designs mostly uses flexure hinges that pivot about axes that lie in the $xy$-plane, taking advantage of the anisotropic resolution of typical 3D printers which extrude a bead of plastic that is typically 2--4 times thinner in height than it is in width. However, the Block Stage is more mechanically complex than the microscope as it has a single moving platform, rather than one platform moving in $xy$ and a second that moves in $z$. This requires levers to provide mechanical reduction and to isolate the motor vibrations from the moving platform. The fact that the levers are actuated relative to the casing of the stage---extruded at the same time from the same material---means that thermal expansion largely cancels out.

\subsection{Lever system}
The moving stage is designed as a table-like structure, with four legs at the corners and three shelves (See \rfig{fig:mechanism}). One shelf situated at the bottom of the legs, a second at the top. A central shelf, $10\mm$ below the top of the legs, is connected to the $z$-axis mechanism and held static in $x$ and $y$. The bottom shelf is actuated in $x$ and $y$ and the top shelf forms the moving platform.  The ratio of the lengths of the top and bottom parts of the legs can adjust the mechanical reduction of $xy$-motion, while the height of the top part determines the travel in the $xy$-plane.  The bottom of the stage is actuated via ``push sticks'' connected to the actuating levers.

The $z$-mechanism consists of two short, wide levers that connect the central shelf of the $xy$-table to the casing of the stage.  The central shelf is extended to the bottom of the stage to allow a vertical separation between the two levers, forming a four bar linkage that allows 1D translation in $z$ without rotation.  The lower lever in the four bar linkage has a vertical actuating lever, connected to the $z$ actuator via a push-stick.  As with the $xy$-table, the actuating lever and the levers responsible for motion need not be the same length, allowing for mechanical reduction.

In all three axes, the relatively simple four-bar mechanisms ensure no rotational motion occurs, but the stage describes an arc rather than a straight line. As such, the moving stage has parasitic motion in $z$ as $x$ and $y$ are adjusted, and similarly there is parasitic motion the $xy$-plane as $z$ is adjusted.  However, the flexure hinges are only bent to a maximum angle of about $\alpha=6\degrees$. The intended motion scales as $\sin\alpha$ and the parasitic motion as $\cos\alpha$, therefore the total parasitic motion when moving from the centre of the stage to the edge of travel is only 50$\um$.

\begin{figure}[t!]
\centering
  \includegraphics[width=.6\textwidth]{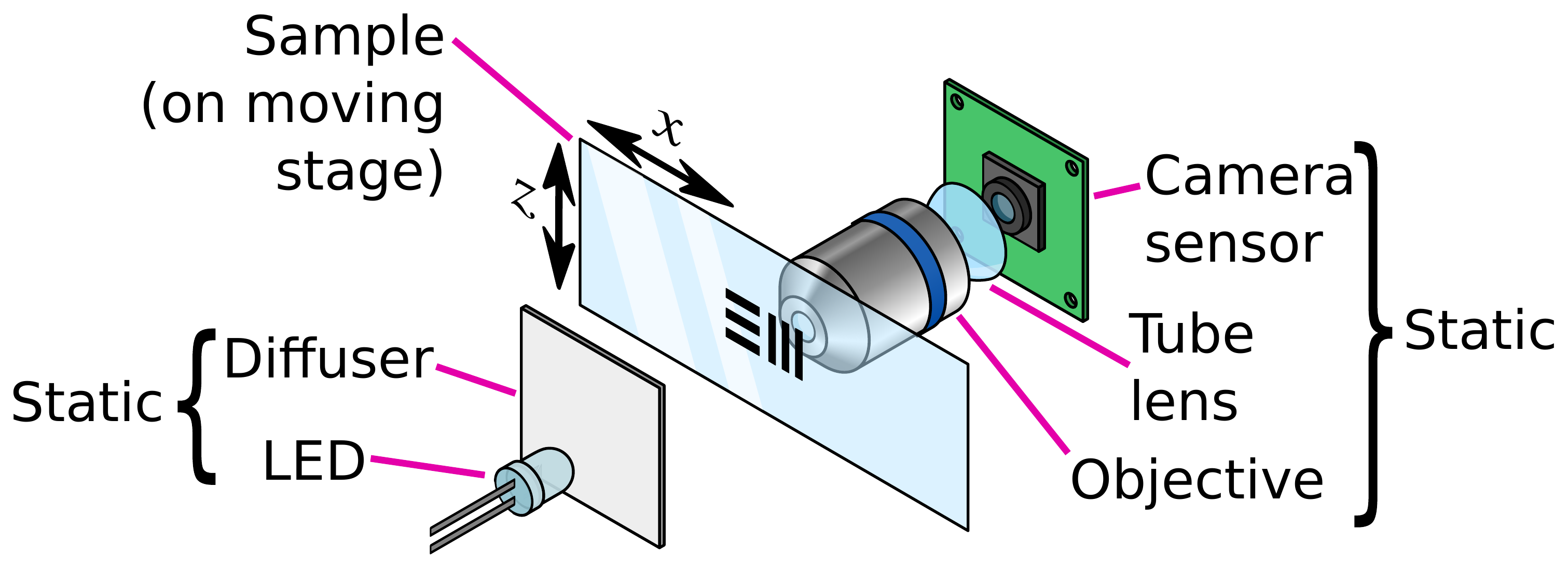}
 \caption{Schematic of the set-up used to track the position of the stage during mechanical characterisation experiments. A diffused LED illuminates a sample mounted on the moving platform. The motion of the sample is tracked by imaging the sample with a microscope optics. Both the LED and the microscope optics are mounted to stationary platforms that form part of the casing of the stage.}\label{fig:TrackPosition}
\end{figure}

\subsection{Actuators and motors}
Each of the three actuators for the stage has a captive brass nut that is driven by the rotation of a steel M3 screw. The actuator design is shared with the Openflexure Microscope, ensuring its assembly, use, and performance are well tested.  Each screw is embedded in a printed control gear that sits on the top of the actuator housing, separated by two washers to reduce wear and ensure smooth rotation. The actuators are tensioned by Viton\textsuperscript{\textregistered} O-rings to ensure that the gear is held firmly against the casing, and to keep the screw and nut under tension to reduce backlash.  

Inexpensive micro-geared stepper motors (28BYJ-48) are used to drive the actuators via printed gears attached to the motor output shaft, these mesh with the stage control gears. The gearing introduces some backlash, though it is no worse than the backlash from the internal gearing of the motor. The stepper motors have 64 steps (technically half-steps) per revolution, and are geared down internally by a factor of 64. The stage is geared down a further factor of 2 by the printed gears to give 8192 steps of the motor per revolution of the control gear. Combined with the $5:1$ mechanical reduction from the levers, each step should thus correspond to a translation of the moving platform by $12.2\nm$. This is comparable to the $10\nm$ expected motion for the smallest microstep of the Newport PZA12 actuators (recommended for Newport fibre alignment stages)\cite{NanoPZManual}.

\subsection{ Drive electronics}

The 28BYJ-48 stepper motors, being unipolar, can be driven very simply using 4 Darlington pairs per motor. We use an Arduino Nano combined with two ULN2003 driver ICs and a small number of passive components to provide a simple motor controller\cite{MotorControllerRepo}. The Arduino controller provides a USB-serial interface between the stage and a host computer, and allows the stage to remember its setting and open loop position even when powered off.

The \verb1openflexure_stage1 Python module provides a high-level interface to the stage, including software backlash correction and error handling.  We do not use micro-stepping, because mechanical reduction of the motor means that there are already sufficiently many steps per revolution of the output shaft.  Micro-stepping also requires more sophisticated motor control, and often introduces vibration even when the stage is stationary, due to the use of pulse width modulation control.

Mitigation against mechanical backlash is provided in the \verb1openflexure_stage1\ Python module by a simple algorithm, which ensures that the stage always approaches its intended position from the same direction. This means that moves in the positive direction are performed as-is, in the negative direction the stage deliberately overshoots before returning to the desired position. The overshoot distance in this work is set to 256 steps of the motor, which we found to be sufficient. While our controller can vary acceleration we have disabled acceleration control for this work.  We opted to run the motors at constant speed instead, as the induced vibrations when accelerating through low speeds had greater effect than the actuator jerk.

\section{Mechanical characterisation}\label{Sec:MechChar}

We performed a series of experiments to assess the mechanical performance of the OpenFlexure Block Stage. The characterisation experiments focused on monitoring the motion of the moving platform relative to the stationary platform. For this, a sample was mounted on the moving platform of the stage, backlit by a diffused LED (see \rfig{fig:TrackPosition}). The sample was then imaged by the optics module of an OpenFlexure Microscope, which consists of a finite-conjugate microscope objective, an achromatic doublet lens ($f=50\mm$, ThorLabs AC127-050-A), and Raspberry Pi camera (Sony IMX219 8MP image sensor). A modified stage with fixed platforms either side of the moving platform (but using an identical mechanism) was used so that both the illumination and imaging optics could be mounted directly to the stage body. The choice of sample and objective for each experiment was dependent on the field of view required. Motion of the sample in the $xz$-plane was tracked by cross-correlating the measured images with a template image taken at the start of the experiment. We calibrated the effective pixel size for images recorded in each experiment by analysing an image of a 1951 USAF resolution test chart (henceforth USAF target; ThorLabs R3L1S4P) taken with the same objective and camera settings as used in the experiment. All characterisation is performed in the $xz$-plane only (except tilt measurements where the $xyz$ volume was scanned), but due to symmetry we expect the $y$-axis to perform similarly to the $x$-axis.

\subsection{Drift}
\begin{figure}[t!]
\centering
\includegraphics[width=\textwidth]{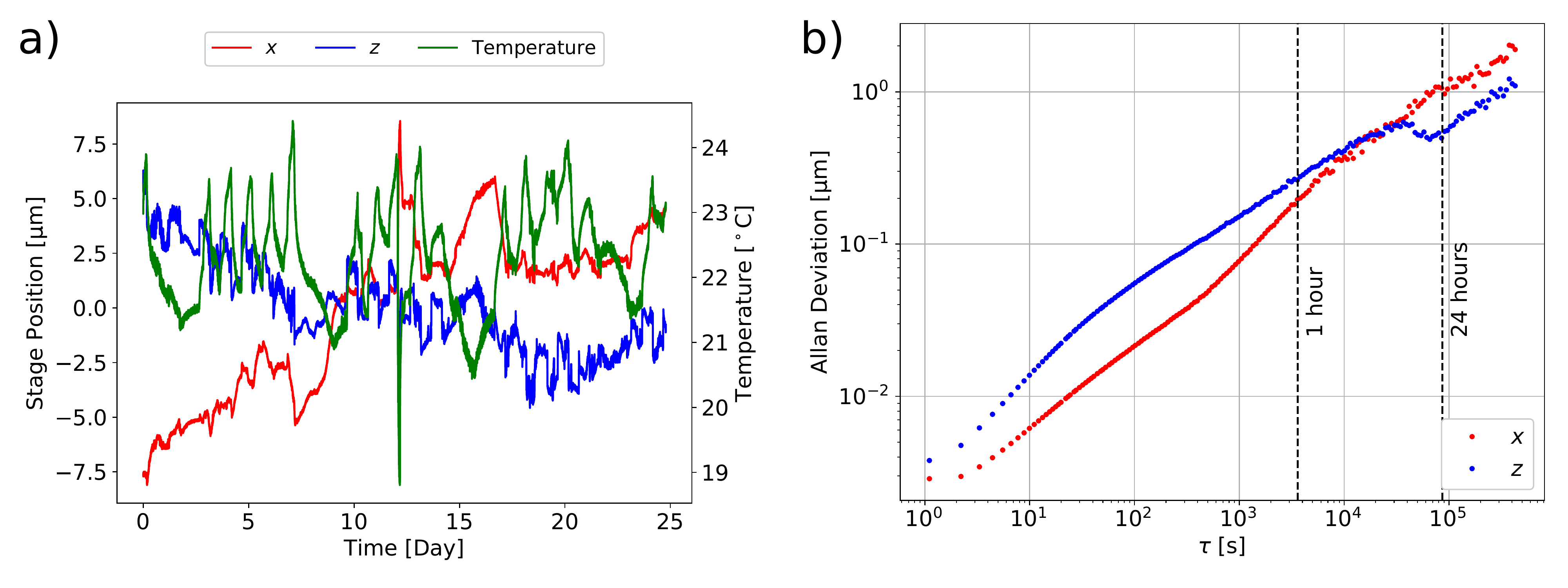}
\caption{a) Drift of the stage over 25 days in ambient conditions. The position of the stage in the $xz$-plane was measured by tracking the position of a scratched microscope slide on the moving platform. b) The Allan deviation of the same data.}
\label{fig:drift}
\end{figure}

Maintaining a fixed position between two objects is critical for applications such as fibre alignment. As such the drift of the moving stage, relative to the fixed stage, was measured over 25 days in ambient conditions. The stage position was measured as described in \rsec{Sec:MechChar} using a scratched microscope slide as a sample and a $10\times$ objective. The temperature of the laboratory was recorded simultaneously. During the experiment the motors were not energised. The mean drift was $0.34\um$ over one hour and $1.6\um$ over a 16 hour period (See \rfig{fig:drift}a). This is comparable to the typical $1.5\um$ drift over 16 hours reported for the Newport ULTRAlign\texttrademark{} fibre stages\cite{ULTRAlign_aplication_note}. The maximum drifts we record are below $7\um$ an hour and $9\um$ in a day, these appear to have been caused by a sudden temperature change in the laboratory.

To further characterise the drift over different time scales we calculate the Allan deviation of the $x$ and $z$-position. Allan Deviation, used widely for measuring oscillator frequency stability, is a statistical measure of deviation over a characteristic time period $\tau$. For a dataset $x(t)$, the data is separated into $M$ blocks of length $\tau$ and the mean of these blocks is calculated
\begin{equation}
\bar{x}_i = \int_{(i-1)\tau}^{i\tau} x(t)\,\mathrm{d}t\,.
\end{equation}
The Allan deviation can then be calculated for a given $\tau$ as
\begin{equation}
\sigma_\mathrm{A}(\tau) = \sqrt{ \frac{1}{2(M-1)}\displaystyle\sum_{i=1}^{M-1} ( \bar{x}_{i+1} - \bar{x}_i )^2 }\,.
\end{equation}

The Allan deviation of the $x$ and $z$-position is shown in \rfig{fig:drift}b. Jitters in processing times meant that the time between adjacent data points varied with a standard deviation of $0.01\,$s, but as the calculation requires a constant data acquisition rate, we used the average rate of 0.9 Hz in our calculations. As the gradient of the Allan deviation is approximately 0.5 in log-log space, this implies that the dominant process is a random walk for all measured time-scales. As the Allan deviation of $z$ is not entirely monotonic this may imply that the stage is reaching the bounds of its random walk, but as the dip corresponds to $\tau=24\,$hours it is more likely related to the stage's response to diurnal temperature variations.

\subsection{Step size, orthogonality, and resolution}
\begin{figure}[t!]
\centering
\includegraphics[width=.9\textwidth]{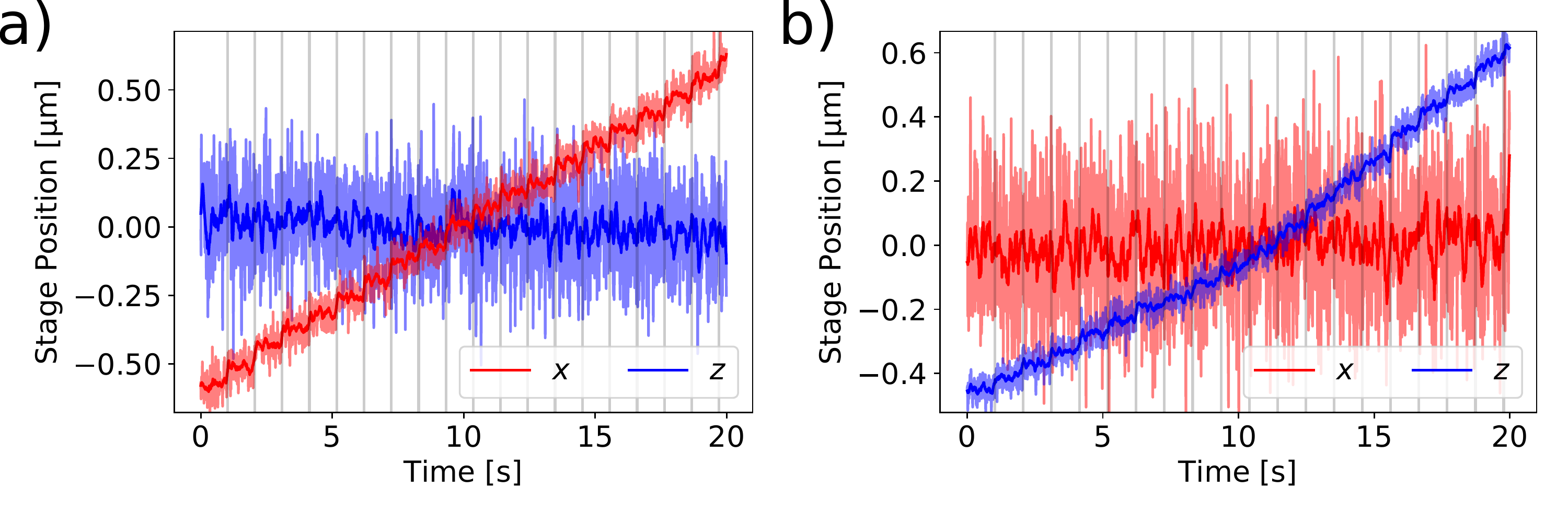}
\caption{The minimum resolvable movement of the stage in the $x$-direction (a) and $z$-direction (b). The solid lines represent an 11-point running average of the data. Vertical grey lines note the time when steps were taken. a) and b) are characteristic examples of the data collected for $x$ and $z$ motion, with measured step sizes of 62\nm (5 steps) and 57\nm (8 steps) respectively.}
\label{fig:resolution}
\end{figure}
The theoretical motion per step is $12.2\nm$ in all axes, however a number of factors from the effective pivot point of the flexures to unwanted compliance in the mechanism can affect the step size. To determine the step size the stage position was measured as the sample was raster scanned over a $10^{5}\times10^{5}$ step$^2$ ($\sim12\times12\mm^2$) area with points that are 5000 steps apart ($\sim60\um$) in the $xz$-plane. The stage position was measured as described in \rsec{Sec:MechChar}, using a USAF target as a sample and a $4\times$ objective.

We calculated an affine transformation matrix, to transform from the position recorded by the motor driver to the measured stage coordinates, by least squares fitting. The affine transformation matrix---an augmented affine matrix---was used to allow for translation, even though the mean position was subtracted from both coordinate systems before fitting. By applying the transformation to unit vectors in the coordinate system of the motors we were able to calculate the mean step size in the $x$ and $z$ directions to be $12.4\pm0.2\nm$ and $9.8\pm0.3\nm$ respectively. The uncertainty was calculated as the standard deviation of the step size from a number of affine transformation matrices fitted to data from regions of the original dataset. The orthogonality of the $x$ and $z$ axes varies from $89.8^\circ$ to $90.8^\circ$ over the area under test. This deviation from orthogonality is expected as the ``push-sticks'' that drive each axis, tilt as the other axes are moved.

Further analysis of the step size (using the data analysed in Section \rsec{Sec:Repeatability}) showed larger relative variations ($\sim20\%$) for moves below 100\um. As this corresponds to one rotation of the screw this implies that the drive screw may not be well aligned to the actuator, but this periodic non-linearity in negligible for larger moves.

As all mechanisms exhibit some level of slip-stick, we also measured the smallest resolvable step the stage can generate. The same set up was used as for measuring the step size, except that the objective was switched to a $40\times$. The stage was moved 10 steps at a time in the $x$ direction with $1\,$s pauses between for $20\,$s. This was repeated with a decreasing number of steps per move until individual movements could no longer be resolved. The resolution was approximately $60\nm$ for each axis, we expect this limit was dominated by measurement noise (See \rfig{fig:resolution}). We note that the measurement noise was significantly higher for the stationary axis($\sigma\sim200\nm$), this may be due to a lack of holding torque applied to this motor.

\subsection{Repeatability}\label{Sec:Repeatability}
\begin{figure}[t!]
\centering
\includegraphics[width=.6\textwidth]{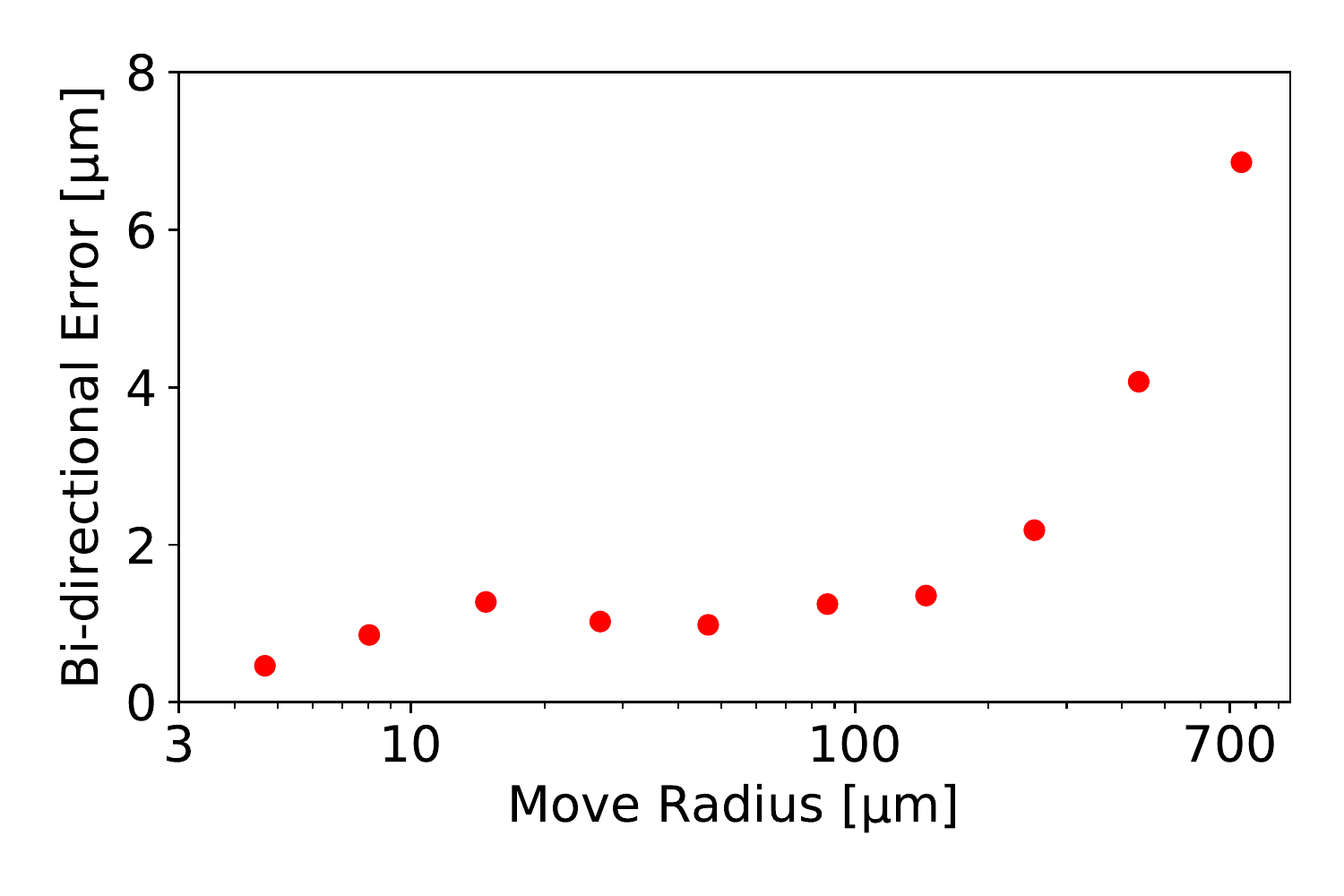}
\caption{Each point represents the mean absolute displacement after twenty bidirectional movements in random directions in the $xz$-plane.}
\label{fig:repeat}
\end{figure}

Repeatability, a measure of how well the stage can return to a position after moving to another point, is critical for applications such as coupling light into an optical fibre. We quantified the bi-directional repeatability error by moving from a given location by a set distance in a random direction on the $xz$-plane and then returning to the original position. The distance moved was varied 2$^{9}$ steps ($\sim5\um$) to 2$^{16}$ steps ($\sim790\um$), with 20 repeats at each distance. The experiment was performed with backlash correction enabled. The stage position was measured as described in \rsec{Sec:MechChar}, using a USAF target as a sample and a $4\times$ objective. The mean absolute displacement after bi-directional moves is shown in \rfig{fig:repeat}. We also note that at small displacements the stage is considerably more repeatable in $x$ than in $z$. This is most likely due to the more complex $z$-mechanism.

\subsection{Tilt}\label{Sec:tilt}

\begin{figure}[t!]
\centering
\includegraphics[width=.9\textwidth]{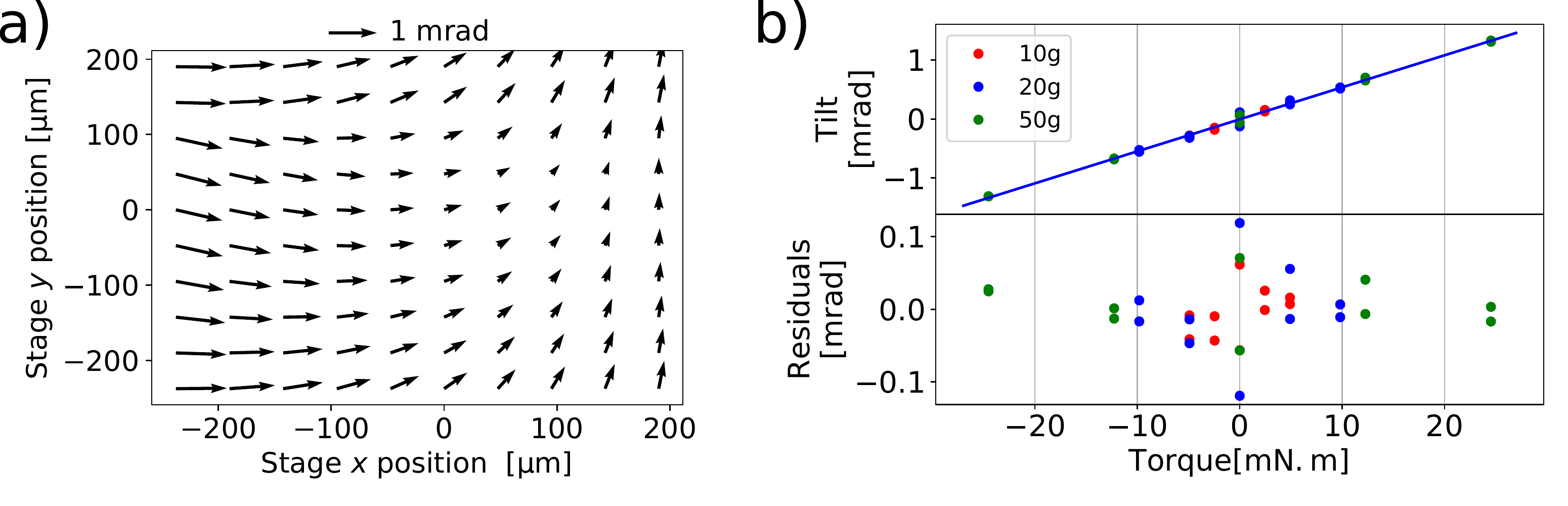}
\caption{a) Parasitic tilt of the moving platform as it is moved in the $xy$-plane. The direction of the arrow notes the direction of the axis of rotation. b) Tilt of the stage in reaction to different torques applied, and the residuals of a linear fit.}
\label{fig:tilt}
\end{figure}

The four-bar mechanisms used to actuate the stage ideally keeps the platform from tilting when translated. However, imperfection in printing and the compliance of nominally rigid portions of the mechanism may lead to parasitic tilt of the moving platform as it is actuated. To measure parasitic tilt the stage was scanned over a $50,000\times50,000\times50,000$ step$^3$ ($\sim0.6\times0.6\mm\times0.6\mm^3$) volume with points that are 5000 steps apart ($\sim60\um$). The tilt about the $x$ and $z$ axes was measured by a custom laser autocollimator which was calibrated against a ThorLabs KMS/M mirror mount. More details on the construction of the laser autocollimator and the calibration are provided in the data archive\cite{Archive}. The parasitic tilt for movements in the $xy$-plane is shown in \rfig{fig:tilt}b), for this section the dominant tilt is around the $x$-axis as the stage moves in the $x$-direction. We notice that in some regions more abrupt changes in tilt are recorded, with the maximum recorded tilt being 3 mrad. The full data set is provided in the data archive\cite{Archive}.

We also measured the tilt of the stage in reaction to an applied torque about the $x$-axis. The torque was applied by mounting a cantilevered structure to the moving platform of the stage. 10 g, 20 g, and 50 g masses were placed on the structure at a fixed displacements from the centre of the platform. The resulting tilt was measured with the laser autocollimator, as shown in \rfig{fig:tilt}b), indicating that the stage has a rotational stiffness of order 18 Nm/rad about the $x$-axis.

\section{Automatic fibre launch alignment}

Coupling a laser beam into an optical fibre is one of the most common micro-positioning tasks in an optics laboratory. This is a challenging task for a positioning stage as displacing the fibre only a few hundred nanometres can lead to a significant drop in efficiency. Aligning single mode fibre by hand can take anything from fifteen minutes to over an hour, and the eventual efficiency is often limited by the operator's skill and patience as well as the experimental system. Here we demonstrate that the OpenFlexure Block Stage can be used to automate this task.
 
For the fibre alignment we focused a 6 mW HeNe laser using a 20$\times$ objective lens onto the end facet of a single mode optical fibre (Thorlabs P1-460B-FC-1) mounted to the moving platform of the stage. The transmitted power was monitored using a TSL2591 high dynamic range light sensor. Before starting automatic alignment the system is roughly aligned by hand.

The fibre alignment algorithm is split in two stages, a coarse alignment algorithm to find an initial faint signal, followed by a gradient ascent algorithm. For coarse alignment the stage is moved in a square spiral in the $xz$-plane ($y$ is the optical axis) with the light sensor set to its highest gain until a power above the background is recorded.

The gradient ascent algorithm is performed on each axis independently. For each axis, the fibre is scanned through 9 equally spaced points, the transmitted power is recorded and a parabola is fitted to the data. If the parabola has a maximum within the measurement range, the stage is moved to this position. Otherwise, a linear fit to the points determines which direction the stage should move to increase intensity, then the algorithm takes another series of nine points. This is repeated until a maximum is found. This process is first performed along the $x$ and $z$ with 400 steps ($\sim4.8\um$) between each point followed by the $y$-axis with 2000 steps ($\sim24\um$) between each point. Once this is completed the number of steps between each point is halved and the process is repeated until the separation between points is 25 steps ($\sim300\nm$). We note that as the final position is the maximum of the parabola, we expect the alignment to be better than the $300\nm$ resolution of the experiment.

\begin{figure}
\centering
\includegraphics[width=.8\textwidth]{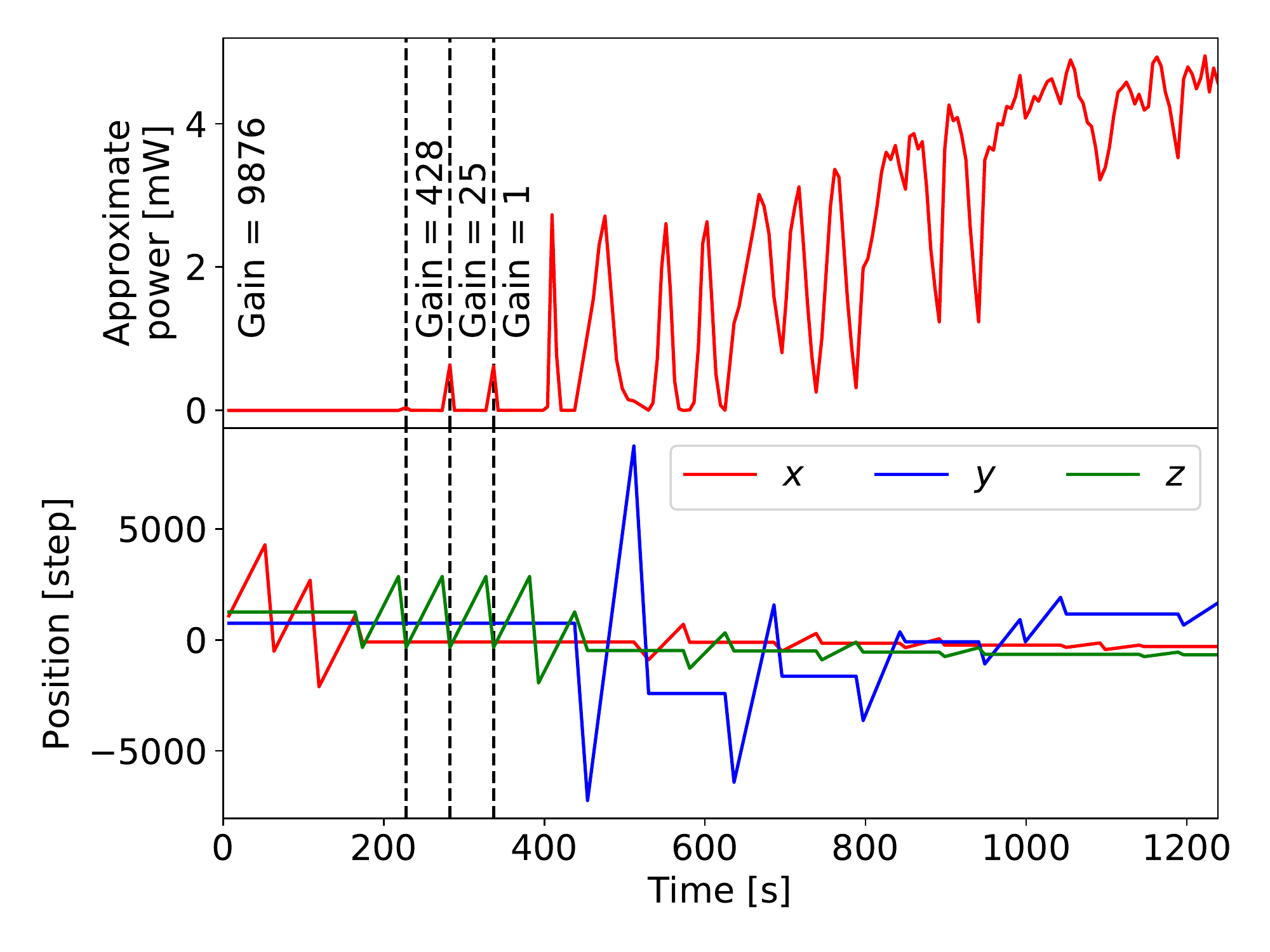}
\caption{An example fibre alignment run performed with our gradient ascent algorithm, showing the sequential alignments increasing the transmitted power. Note that some measurements saturated the power sensor and had to be repeated with a lower gain.}
\label{fig:alignment}
\end{figure}

The data recorded during a gradient ascent is shown in \rfig{fig:alignment}. Over the course of about 20 minutes, the signal coupled into the single mode fibre increased from a few hundred nW to over $4\,$mW. Over multiple runs, we found that the success rate of the gradient ascent algorithm increased if the threshold of the coarse alignment was set higher as this lowered the chances of getting stuck at a local maximum. We note that the laser powers presented are approximate due to issues coupling the light onto TSL2591, with $\sim30\%$ variation in signal when connecting the FC connector to our housing for the sensor. This does not affect the performance of the alignment algorithm as FC connector is not disconnected during an alignment run.

\section{Conclusion}

The OpenFlexure Block Stage demonstrates that 3D printed flexure mechanisms can be used to produce a high precision translation stage. The stage has sub-100nm resolution and a $2\times2\times2\,\mm^3$ travel range. Due to its monolithic design its mean drift over a 16 hour period is only $1.6\um$. The stage can be motorised with readily available stepper motors, allowing automated or remote control of experiments. We have demonstrated that the OpenFlexure Block Stage can perform fully automated fibre coupling making it a viable alternative to commercially available auto-aligners that are usually two orders of magnitude more expensive. Automated fibre alignment not only saves time, but could be implemented to periodically realign long running experiments without the need for an operator to be present.

\section{Acknowledgements}

The authors wish to thank Abhishek Ambekar and Roddy Jaques for their contribution to early versions of data collection scripts, and Joel T. Collins for help with producing Figures 1 and 3.

The authors would like to acknowledge financial support from EPSRC (EP/P029426/1), the Royal Commission for the Exhibition of 1851, and the Royal Society (URF\textbackslash{}R1\textbackslash{}180153, RGF\textbackslash{}EA\textbackslash{}181034).

The data collection scripts, analysis scripts, raw data, and hardware design files used for this paper are available at \cite{Archive}.


\begin{thebibliography}{10}

\bibitem{silver2019five}
A.~Silver,
\newblock Nature {\bf 565}, 123 (2019).

\bibitem{RepRap_bowyer_2011}
R.~Jones {\em et~al.},
\newblock Robotica {\bf 29}, 177–191 (2011).

\bibitem{kamthai2015thermal}
S.~Kamthai and R.~Magaraphan,
\newblock Thermal and mechanical properties of polylactic acid (pla) and
  bagasse carboxymethyl cellulose (cmcb) composite by adding isosorbide
  diesters,
\newblock in {\em AIP Conference Proceedings} Vol. 1664, p. 060006, AIP
  Publishing, 2015.

\bibitem{CRC_MechProps}
J.~R. Rumble, editor,
\newblock {\em CRC Handbook of Chemistry and Physics, 100th Edition} (CRC
  Press/Taylor \& Francis, 2019), .

\bibitem{riddell1966fatigue}
M.~Riddell, G.~Koo, and J.~O'Toole,
\newblock Polymer Engineering \& Science {\bf 6}, 363 (1966).

\bibitem{BlockstageRepo}
{GitLab repository of the OpenFlexure Block Stage
  \url{https://gitlab.com/openflexure/openflexure-block-stage}}.

\bibitem{sharkey2016one}
J.~P. Sharkey, D.~C. Foo, A.~Kabla, J.~J. Baumberg, and R.~W. Bowman,
\newblock Review of Scientific Instruments {\bf 87}, 025104 (2016).

\bibitem{ofm2019}
J.~T. Collins {\em et~al.},
\newblock {Robotic microscopy for everyone: the OpenFlexure Microscope {\it
  Submitted} 2019}.

\bibitem{NanoPZManual}
{Newport Corporation},
\newblock {NanoPZ\texttrademark{} User Manual, 90043104 Rev. A}.

\bibitem{MotorControllerRepo}
{GitHub repository of Motor controller hardware, firmware and software
  \url{https://github.com/rwb27/openflexure_nano_motor_controller}}. A version
  of this repository is archived with the dataset for this paper.

\bibitem{ULTRAlign_aplication_note}
{Newport Corporation},
\newblock {Application Note: ULTRAlign\texttrademark{} Precision Fiber Optic
  Positioning System
  \url{https://www.newport.com/n/ultralign-precision-fiber-optic-positioning-system}}.

\bibitem{Archive}
{Meng, Q., Harrington, K., Stirling, J., Bowman, R., in press. Dataset for "The
  OpenFlexure Block Stage: Sub-100 nm fibre alignment with a monolithic plastic
  flexure stage". Bath: University of Bath Research Data Archive.
  \url{https://doi.org/10.15125/BATH-00737}}.

\end{thebibliography}
\end{document}